 \documentclass[final,3p,times,twocolumn]{elsarticle}
 
\usepackage{amssymb}

\usepackage{ulem}
\usepackage{color}

\begin{document}

\begin{frontmatter}

\title{Heavy quark \& direct photon production and heavy quark parton densities}

%
\author{T. Stavreva}
\ead{stavreva@lpsc.in2p3.fr}
\address{Laboratoire de Physique Subatomique et de Cosmologie, UJF, CNRS/IN2P3,
\\
INPG, 53 avenue des Martyrs, 38026 Grenoble, France}

%





\begin{abstract}
Direct photon production in association with a heavy quark can serve as an excellent 
tool for the study of the heavy quark distributions.  Currently it is believed 
that heavy quarks are produced radiatively inside the nucleus, and so there is no 
need to take into account heavy quark parameters inside global PDF analysis.  
Certain models taking into account the possibility of an intrinsic charm component 
exist.  Here we present how these affect the $\gamma + c$ cross section. While at $pA$ collisions 
the potential of this process to constrain the gluon nuclear PDF which is currently largely unconstrained 
is presented. 
\end{abstract}


\end{frontmatter}

\section{Introduction}
\label{Introduction}

\subsection{Direct $\gamma + Q$}
In order to gather more information on the heavy quark PDFs it is important to investigate in detail processes 
sensitive to them.  
Direct photon production in association with a heavy quark (c or b) is exactly one such process due to its 
dependence on subprocesses sensitive to the initial state heavy quark (HQ) distributions  \cite{Stavreva:2009vi}.  

At LO  it has a  simple form -- 
it consists of one hard-scattering subprocess, $g+Q\rightarrow \gamma+Q$ (Compton subprocess), 
and also fragmentation contributions\footnote{The LO fragmentation contributions consist of 
all $2\rightarrow 2$ subprocesses of order ${\cal O} (\alpha_s^2)$ containing at least one 
heavy quark in the final state convoluted with the photon fragmentation function 
$D_{\gamma/q,g}(z,Q^2)$.}, these being greatly suppressed due to experimental isolation requirements.
At NLO the number of contributing hard-scattering subprocesses increases to seven {\footnote {This 
calculation is performed in the variable flavor number scheme, where the heavy quarks are treated as massless.}}:
\vspace{-0.6cm}
\begin{center}
 \begin{tabular}{p{1.7in}lp{1.7in}l} 
  $g+g\rightarrow Q+\bar Q+\gamma$& $Q+Q\rightarrow Q+Q+\gamma$\\ 
  $g+Q\rightarrow g+Q+\gamma$& $Q+\bar Q\rightarrow Q+\bar Q+\gamma$\\ 
  $Q+q\rightarrow q+Q+\gamma$& $q+\bar q\rightarrow Q+\bar Q+\gamma$ \\ 
  $Q+\bar q\rightarrow Q+\bar q+\gamma$\\  
 \end{tabular}
\end{center}
also including NLO fragmentation effects, which are small due to the isolation cuts as in the LO case. 
Given that most of the subprocesses are initiated by either heavy quarks or gluons, and in the proton the gluon distribution is well known, it is possible to focus on constraining the heavy quark PDF through measurements of this process. 

\subsection{Intrinsic Charm}

\begin{figure}[!h]
	\centering
	\includegraphics[width=0.8\columnwidth]{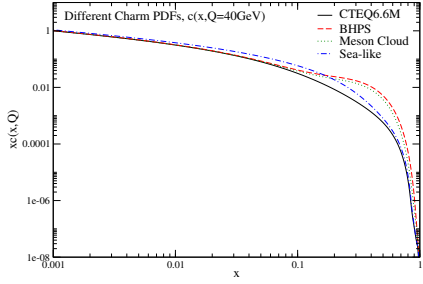}
	\caption{Comparison between the radiatively generated charm PDF (CTEQ6.6M, solid black line), 
                 charm PDF including IC: under the BHPS model (CTEQ6.6C1, dashed red line), 
		 under the Meson Cloud model (CTEQ6.5C4, dotted green line), 
                 under the sea-like model (CTEQ6.6C3, dash-dotted blue line)}
	\label{fig:IC}
\end{figure}

In the standard global analysis of PDFs, heavy quarks do not have free fit parameters associated with them. 
Therefore currently it is assumed that there is no intrinsic charm (IC) or intrinsic bottom (IB) inside 
the nucleus.  
This means that the heavy quark PDF can be obtained purely perturbatively through
the use of the DGLAP evolution equations, having set an appropriate initial condition 
(such as $f_Q(x,Q_0=m_Q)=0$).  
However there is no theoretical reason that this should be the case and moreover there have 
been some data that point towards the existence of an intrinsic charm
component in the proton ($F_2^c$ at large $x$ as measured by EMC \cite{Aubert:1982tt}). 
As a result there are non-perturbative models predicting the size and shape of the intrinsic heavy quark component
\cite{Brodsky:1980pb, Navarra:1995rq} which are used in PDF global fits \cite{Pumplin:2007wg} \footnote{Currently 
there are fits available only for IC.  The IB/IC ratio is expected to be proportional to $m_c^2/m_b^2$.}. 
In Fig. \ref{fig:IC} the difference between different IC PDFs  and the radiatively radiate charm (CTEQ6.6M) 
is displayed.  
Both the BHPS (dashed red line) and Meson Cloud model (dotted green line) display a similar behavior where they peak at high $x$, while being very similar to the radiatively generated charm (solid black line) at lower $x$.  
The purely phenomenologically inspired sea-like model (dash-dotted blue line) produces a PDF which is larger than the radiatively generated one in a broad $x$ range.  
In order to constrain precisely the HQ PDF it is important 
to measure processes sensitive to it such as $\gamma + Q$.  
How these different models affect 
the $\gamma + Q$ cross-section is presented in the next section.
\vspace{-0.3cm}
\section{Predictions: hadron-hadron collisions}
\subsection{Tevatron}
\begin{figure}[!h]
	\centering
	\includegraphics[width=0.8\columnwidth]{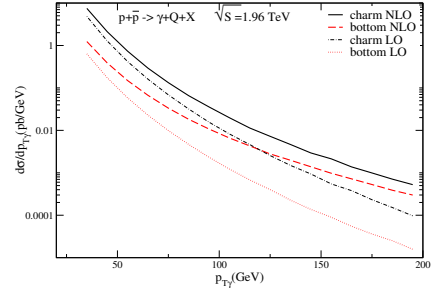}
	\caption{$d\sigma^{\gamma + Q}/dp_{T\gamma}$: for charm at NLO (black solid line), at LO (black dash-dotted line), for bottom at NLO (red dashed line), at LO (red dotted line).}
	\label{fig:Tevatron1}
\end{figure}

\begin{figure}[!h]
	\centering
	\includegraphics[width=0.8\columnwidth]{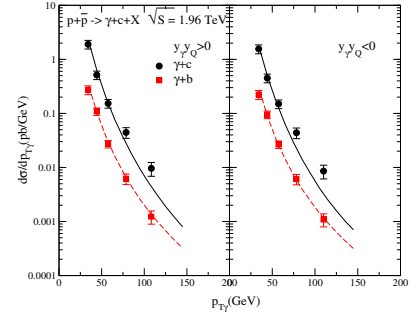}
	\caption{Comparison between theory - $\gamma+c$ (solid black curve), $\gamma+b$ (dashed red curve) and D\O\ data -
$\gamma+c$ (black circles), $\gamma+b$ (red squares).}
	\label{fig:D01}
\end{figure}

\begin{figure}[!h]
	\centering
	\includegraphics[width=0.8\columnwidth,trim=0cm 1.5cm 0cm 0cm,clip]{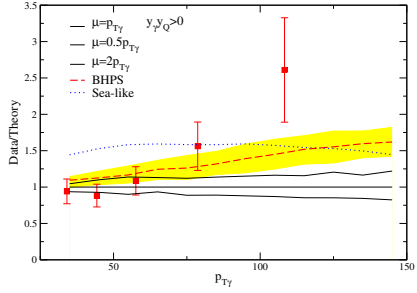}
	\caption{Data over theory ratio for $\gamma +c$ production (red squares), the theoretical predictions are for radiatively generated charm including the scale dependence (black lines), BHPS IC (red dashed line), sea-like IC (blue dotted line).}
	\label{fig:D02}
\end{figure}

The differential cross-section vs $p_{T\gamma}$ for $\gamma+c$ and $\gamma+b$ production 
at the Tevatron is presented at both NLO and LO in Fig.\ \ref{fig:Tevatron1}. 
From there we see that the difference between  $d\sigma^{\gamma+c}/dp_{T\gamma}$  and 
$d\sigma^{\gamma+b}/dp_{T\gamma}$ at LO stays almost constant, whereas it decreases 
at NLO with growing $p_{T\gamma}$.  
The reason for this is that at the Tevatron, 
due to the abundance of valance quarks and anti-quarks, the annihilation subprocess 
($q+\bar q\rightarrow Q+\bar Q+\gamma$) starts to dominate the cross-section, and this 
subprocess is exactly the same for both $\gamma+c$ and $\gamma+b$ production. 
When one looks at the comparison between theory and experimental measurements (performed 
by the D\O\ collaboration \cite{Abazov:2009de} ) in Fig.\ \ref{fig:D01}, the agreement for $\gamma+b$ is 
very good, whereas for $\gamma+c$ the data overshoot the theoretical predictions at high $p_T$. 
One possibility to try and correct for this discrepancy is to utilize IC PDFs in the theoretical predictions. 
In Fig.\ \ref{fig:D02} the data over theory ratio is presented, as well as the ratios 
${d\sigma^{BHPS}/dp_{T\gamma}}\over{d\sigma^{CTEQ6.6M}/dp_{T\gamma}}$, 
${d\sigma^{sea-like}/dp_{T\gamma}}\over{d\sigma^{CTEQ6.6M}/dp_{T\gamma}}$. 
As expected the sea-like cross-section overshoots the data at low $p_T$, but undershoots it 
at high $p_T$.  
Conversely the BHPS cross-section, follows the data trend, and provides 
a better description than both the radiatively generated charm and the sea-like charm 
cross-section, however it still undershoots the data at very high $p_T$. 
Therefore in order to confirm or disprove the existence of IC more data for 
processes sensitive to IC is needed.   
In the next subsection the effects of IC on the 
direct photon and heavy quark cross-section, and the possibility of constraining IC 
at the LHC are presented.  
\vspace{-0.2cm}
\vspace{-0.2cm}
\vspace{-0.5cm}
\subsection{LHC} 
\begin{figure}[!h]
	\centering
	\includegraphics[width=0.8\columnwidth,trim=0cm 1.5cm 0cm 0cm,clip]{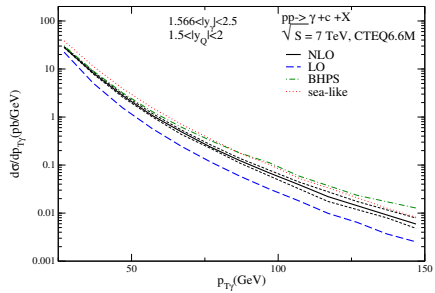}
	\caption{$d\sigma^{\gamma+c}/dp_{T\gamma}$ at CMS for forward rapidity for radiatively generated charm (black solid line) + scale dependence (black dashed curves), BHPS IC (green dash-dotted line), sea-like IC (red dotted line).}
	\label{fig:CMS1}
\end{figure}

\begin{figure}[!h]
	\centering
	\includegraphics[width=0.8\columnwidth,trim=0cm 1.5cm 0cm 0cm,clip]{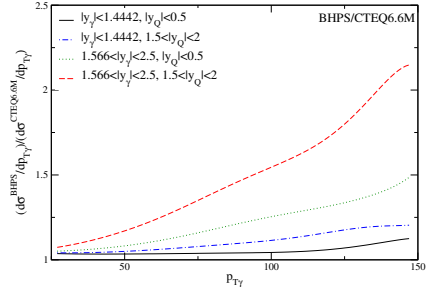}
	\caption{Ratio of BHPS IC $\gamma+c$ cross-section over one using radiatively generated charm for different restrictions on the photon and charm rapidity.}
	\label{fig:CMS2}
\end{figure}

At the LHC, already at $7$ TeV the $x$ probed is small ($x\sim p_T/\sqrt{S}$).  
Therefore in order to be sensitive to the BHPS IC, which shows up at large-$x$ (Fig. \ref{fig:IC}), 
one has to probe forward rapidities in order to be sensitive to it.
Utilizing experimental cuts appropriate for the CMS detector \cite{cms-det1,cms-det2} and focusing on the 
forward rapidity region ($1.566 <|y_\gamma| < 2.5$, $1.5 \le |y_Q| < 2.0$) for both photons and 
c-quarks, we present in Fig.\ \ref{fig:CMS1} the predictions for $d\sigma^{\gamma+c}/dp_{T\gamma}$ at $\sqrt{S}=7$ TeV. 
The BHPS cross-section (green dashed line) is similar to the radiatively generated one (black solid curve) 
at low $p_T$, but as $p_T$ increases (and correspondingly $x$ increases) so does the difference between the 
two curves, thus providing a possibility for testing the existence of BHPS IC. 
To portray the need for measurements at forward rapidities the ratio between the BHPS cross-section and the 
radiatively generated one for several different combinations of the photon 
and heavy quark rapidities is presented in Fig.\ \ref{fig:CMS2}. 
There the ratio at larger $p_T$ increases from $\sim 1.1$ for $|y_\gamma| < 1.4442$ \& $|y_Q| < 0.5$, to up
to $\sim 2.2$ for $1.566 <|y_\gamma| < 2.5$ \& $1.5 \le |y_Q| < 2.0$.  
\section {Predictions: hadron-ion collisions} 
\begin{figure}[!h]
	\centering
	\includegraphics[width=0.8\columnwidth,trim=0cm 1.5cm 0cm 0cm,clip]{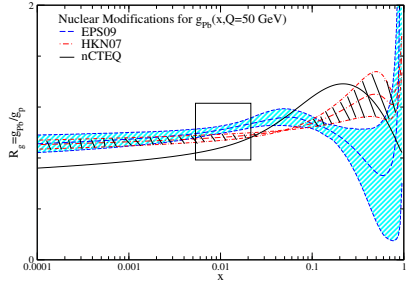}
     	\caption{The gluon nuclear modification factor for nCTEQ (solid black line), HKN07 (red dash-dotted band), 
EPS09 (dashed blue band).}
	\label{fig:nPDF1}
\end{figure}

\begin{figure}[!h]
	\centering
	\includegraphics[width=0.8\columnwidth,trim=0cm 1.5cm 0cm 0cm,clip]{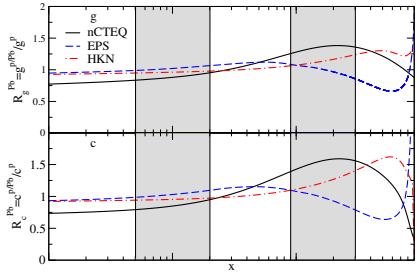}
	\caption{Comparison between the charm and gluon nuclear modification factors for nCTEQ (solid black line), HKN07 (red dash-dotted band), 
EPS09 (dashed blue band), the two shaded regions represent the $x$ probed at LHC (first band) and at RHIC (second band).}
	\label{fig:nPDF2}
\end{figure}

\begin{figure}[!h]
	\centering
	\includegraphics[width=0.8\columnwidth,trim=0cm 1.6cm 0cm 1.05cm,clip]{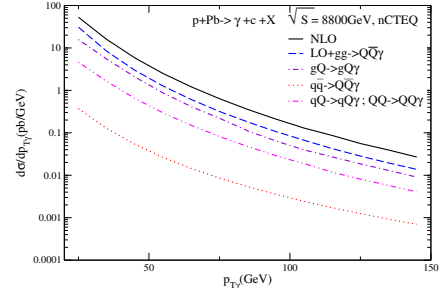}
	\caption{Subprocess contributions to $\sigma^{\gamma+c}$ at p-A collisions at ALICE.}
	\label{fig:ALICE1}
\end{figure}

\begin{figure}[!h]
	\centering
	\includegraphics[width=0.8\columnwidth,trim=0cm 1.6cm 0cm 1.05cm,clip]{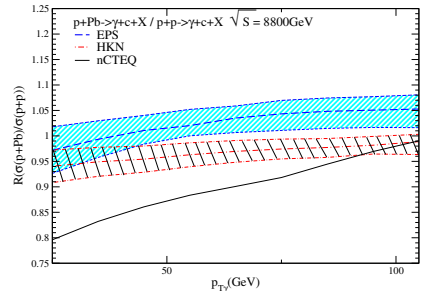}
	\caption{Nuclear modification factor for $\gamma+c$ at ALICE, using nCTEQ (black solid line), EPS09 (blue dashed line) + error band, 
HKN07 (red dash-dotted line) + error band.}
	\label{fig:ALICE2}
\end{figure}

It is also interesting to study the IC effects in $pA$ collisions. 
However in order for such a study to be performed, the gluon nuclear PDF (nPDF) should 
be as well known as in the free proton.  
Unfortunately this is not the case, and the gluon distribution is only very 
weakly constrained by NMC data on $F_2^D(x,Q^2)$ and $F_2^{Sn}/F_2^C(x,Q^2)$ 
\footnote{This is true for most nPDF fits, while EPS09 also includes 
data on $\pi^0$ production at RHIC.}.
%
The large uncertainty in the nuclear gluon as well as the variations in  
predictions of different fits (nCTEQ \cite{Schienbein:2008pw,Schienbein:2009kk,Kovarik:2010uv}, 
HKN07 (red dash-dotted band) \cite{Hirai:2007sx}, 
EPS09 \cite{EPS}) is presented in Fig.\ \ref{fig:nPDF1} in the form of the gluon nuclear 
modification factor $R_g(x,Q)=g^{p/Pb}(x,Q)/g^p(x,Q)$. 
It is quite clear that there is a strong need for data constraining the gluon nPDF. 
In the general framework, where there is no IC considered, and the charm distribution
is solely based on the gluon PDF, the charm nuclear modification, $R_c$, follows $R_g$ 
quite closely as shown in Fig.\ \ref{fig:nPDF2}. 
Since $\sigma^{\gamma + c}$ is quite sensitive to the gluon PDF as well as the charm PDF as shown in Fig.\ \ref{fig:ALICE1}, 
where the dominating subprocesses at $\sqrt{S}=8.8 \rm TeV$  for $p-A$ collisions are 
$gg$ and $gQ$ initiated, it is a very useful process for constraining the gluon nPDF. 
Focusing on the central rapidity region in ALICE will isolate any effects 
caused by the possible presence of the theoretically favored BHPS-IC, which will 
appear at large-$x$ and forward rapidity, Fig.\ \ref{fig:CMS2}.  Therefore we can expect that the nuclear corrections to $\sigma^{\gamma + c}$ will follow the ones for $R_g$ and $R_c$ correspondingly.  
In Fig.\ \ref{fig:ALICE2} the nuclear modification factor, $R^{\sigma^{\gamma+c}}={{d\sigma/dp_{T\gamma}(pA)} \over {A_{Pb}d\sigma/dp_{T\gamma}(pp)}}$, to the direct photon and heavy quark cross-section is shown.  It can be seen clearly by comparing Fig.\ \ref{fig:ALICE2} and the boxed region in 
Fig.\ \ref{fig:nPDF1} that the nuclear modification factor to the cross-section, 
$R^{\sigma^{\gamma+c}}$, follows closely the nuclear modifications for the gluon nPDF ($R_g$), in the region of x probed at the LHC; for more details see \cite{Stavreva:2010mw}.   Therefore this process is an excellent candidate for constraining the gluon nuclear distribution. 
\vspace{-0.5cm}

\vspace{-0.7cm}
\section{Conclusion} 
It was shown that through direct photon production in association with a heavy quark we can study and constrain the heavy quark PDFs.  More 
specifically measurements at the Tevatron of this process have indicated a possible existence of IC that seems to favor the BHPS model.  
Further tests at the LHC  of this process especially at forward rapidities should be able to constrain or disfavor the presence of intrinsic charm 
in the proton.  While such a constrain inside the nucleus is also very important, one has to have a quite precisely determined gluon nPDF 
in order to do so.  Since this is not the case, we have shown here that direct photon + heavy quark production at pA collisions is also 
a great process for constraining the nuclear gluon PDF. 






\end{document}